\documentclass[10pt,conference]{IEEEtran}
\IEEEoverridecommandlockouts

\usepackage{cite}
\usepackage{amsmath,amssymb,amsfonts}
\usepackage{algorithm}
\usepackage{algpseudocode}
\usepackage{xspace}
\usepackage{booktabs}
\usepackage{multirow}
\usepackage{multicol}
\usepackage{enumitem}
\usepackage{graphicx}
\usepackage{textcomp}
\usepackage{listings}
\usepackage[dvipsnames]{xcolor}
\lstdefinelanguage{json}{
  morestring=[b]",%
  morestring=[s]{'}{'},%
  morecomment=[l]{//},%
  morecomment=[s]{/*}{*/},%
  morekeywords={true,false,null},%
  sensitive=false
}
\usepackage{url} 
\def\BibTeX{{\rm B\kern-.05em{\sc i\kern-.025em b}\kern-.08em
T\kern-.1667em\lower.7ex\hbox{E}\kern-.125emX}}

\lstdefinestyle{promptstyle}{
  basicstyle=\ttfamily\scriptsize, %
  breaklines=true,
  breakindent=0pt,
  frame=none,
  columns=fullflexible,
  keepspaces=true,
  showstringspaces=false,
  tabsize=2,
  literate={->}{{$\rightarrow$}}2 {⊥}{{$\bot$}}1
}

\usepackage{bm}
\makeatletter
\AtBeginDocument{\DeclareMathVersion{bold}
\SetSymbolFont{operators}{bold}{T1}{times}{b}{n}
\SetMathAlphabet{\mathrm}{bold}{T1}{times}{b}{n}
\SetMathAlphabet{\mathit}{bold}{T1}{times}{b}{it}
\SetMathAlphabet{\mathbf}{bold}{T1}{times}{b}{n}
\SetMathAlphabet{\mathtt}{bold}{OT1}{pcr}{b}{n}
\SetSymbolFont{symbols}{bold}{OMS}{cmsy}{b}{n}
\renewcommand\boldmath{\@nomath\boldmath\mathversion{bold}}}
\makeatother
\newcommand{\name}{{\sc RAMD}\xspace}
\newcommand{\longName}{Reference Architecture Mapping Discovery\xspace}
\newcommand{\Pall}{\ensuremath{P_{\mathrm{all}}}\xspace}
\newcommand{\Pcomp}{\ensuremath{P_{\mathrm{comp}}}\xspace}
    
\begin{document}

\title{Discovering the Relation Between Architectural Models and Reference Architectures}
\author{
\IEEEauthorblockN{Thi Dinh Tran}
\IEEEauthorblockA{\textit{dept. Computer Science} \\
\textit{Gran Sasso Science Institute}\\
L'Aquila, Italy\\
thidinh.tran@gssi.it}
\and
\IEEEauthorblockN{Maria Teresa Rossi}
\IEEEauthorblockA{\textit{dept. Computer Science} \\
\textit{Gran Sasso Science Institute}\\
L'Aquila, Italy\\
mariateresa.rossi@gssi.it}
\and
\IEEEauthorblockN{Ludovico Iovino}
\IEEEauthorblockA{\textit{dept. Computer Science} \\
\textit{Gran Sasso Science Institute}\\
L'Aquila, Italy\\
ludovico.iovino@gssi.it}
\and
\IEEEauthorblockN{Leonardo Mariani}
\IEEEauthorblockA{\textit{DISCo} \\
\textit{University of Milano-Bicocca}\\
Milan, Italy\\
leonardo.mariani@unimib.it}
}
\maketitle

\begin{abstract} Assessing whether concrete \emph{software architectures} are consistent with their \emph{reference architecture} (RA) is essential to ensure that architectural decisions scale coherently across projects, preserving the quality attributes the RA was designed to guarantee. In the absence of dedicated techniques, mapping the components of an architectural model onto a predefined RA can be addressed only with general-purpose model-comparison algorithms. Our experiments show that these algorithms are inaccurate when applied to RAs. 
To address this gap, we propose \longName (\name), an LLM-based workflow that combines \emph{model analysis} with \emph{structured prompting}, and refines results through an \emph{iterative RA compliance-checking mechanism}.
We evaluate \name on 64 open-source architectural models spanning three RAs from the IoT, self-driving, and smart parking domains. %
With domain knowledge and GPT-4o, \name reaches 85.8\% mapping accuracy and outperforms the strongest baseline by 44.6 percentage points, while ensuring structural consistency with the RA.
\end{abstract}

\begin{IEEEkeywords}
Software Architecture, AADL, Reference Architecture, Component Mapping, LLMs, Architecture Compliance
\end{IEEEkeywords}

\section{Introduction}
\label{sec:introduction}

A \emph{Reference Architecture} (RA) is a domain‑specific, generic architectural blueprint that codifies the structural and functional principles shared by systems belonging to the same family of applications~\cite{angelov2012framework,nakagawa2014reference,cloutier2010concept}. It captures design knowledge that has been validated and refined across multiple projects, providing a stable foundation for developing consistent and interoperable system architectures.

Exploiting RAs to guide the definition of concrete architectural models is crucial for ensuring that systems remain compliant with established domain knowledge and proven design principles, meet key quality attributes (e.g., reliability, security, modifiability) and organization‑wide objectives, and reduce integration and evolution risks by preventing uncontrolled architectural drift~\cite{RAMLI2025112264}.
Major cloud providers actively endorse the use of RAs to disseminate consolidated design knowledge. Amazon Web Services and Google Cloud maintain a catalog of reference architectures grounded in their Well-Architected Frameworks~\cite{aws-architecture,gcp-architecture}. Microsoft likewise stresses its importance, publishing hundreds of reference architectures in the Azure Architecture Center and supports them as proven best practices and patterns~\cite{azure-wellarchitected}.

Once compliance with RAs is established, organizations can exploit it in several ways. At the project level, it supports automated impact analysis when a model evolves~\cite{lehnert2011taxonomy} and enables the systematic reuse of conforming architectural assets~\cite{garces2021three}. At the organizational level, it underpins architectural governance~\cite{de2012controlling} and the continuous monitoring of architectural technical debt~\cite{besker2018managing}.
In this way, reference architectures become not only design blueprints but operational instruments for ensuring consistency, predictability, and long‑term sustainability of software systems.

Many architectural models are now available across domains and repositories~\cite{fuchss2022establishing,lopez2024modelset,glaser2025extended}, yet very little is known about whether these models are actually compliant with their intended reference architectures, leaving organizations with a growing collection of artifacts whose architectural validity is uncertain. This gap is problematic: without explicit compliance knowledge, models cannot reliably support reuse and governance, and they risk drifting away from the quality attributes and constraints encoded in the reference architecture. Automatically determining compliance, however, remains a significant challenge because it requires discovering semantic correspondences between heterogeneous modeling constructs and interpreting architectural intent. In fact, models and RAs are typically authored by different stakeholders, each using their own vocabulary. The same architectural role may appear under different names, granularities, or decompositions, producing both semantic and structural mismatches that lexical or local reasoning cannot resolve. 
As a result, organizations possess many models, but lack the means to understand how these models relate to the available architectural standards.

Architecture conformance techniques, such as reflexion models~\cite{murphy2001reflexion,passos2009static}, presuppose that the mapping between models has already been declared. \emph{Lexical matching} techniques can exploit TF-IDF representation of components~\cite{salton1988termweighting,manning2008introir}, as well as generic schema~\cite{shvaiko2011ontology} to discover relations between models and reference architectures, but cannot enforce the structural and mandatory constraints of an RA. 
Similarly, \emph{Sentence Embedding} can abstract from specific identifiers to discover mappings between components, but fails to address the conformance constraints deriving from reference architectures~\cite{reimers2019sentencebert,yin2024study}. Alternatively, models and reference architectures can be \emph{compared as graphs}, to discover possible matching components and relations~\cite{conte2004thirty,tran2026structuralsemantic}. Yet graphs can hardly capture the semantic relation between models and RAs.

To address these limitations, we propose \longName (\name), an LLM-assisted mapping pipeline that combines model analysis, the retrieval of domain knowledge from previously discovered mappings, structured LLM prompting, and a compliance-guided iterative refinement loop to automatically reconstruct the conformance relation between a model and its reference architecture. Our key insight is that LLMs can bridge vocabulary and granularity gaps when reasoning about individual component pairs, but they cannot easily enforce the global RA constraints that a valid mapping must satisfy. We therefore \emph{couple the LLM with an external compliance checker in a closed iterative loop}: the LLM proposes a candidate mapping, and the checker validates the candidate against the RA constraints, returning any violation to the LLM as feedback for the next iteration.

We further equip \name with \emph{domain knowledge} about mappings between models and reference architectures, exploiting a retrieval-augmented generation (RAG) mechanism~\cite{lewis2020retrieval}. %

We implemented \name for the Architecture Analysis \& Design Language (AADL) model, which is an SAE International standard~\cite{feiler2012model} for the model-based design of safety- and performance-critical embedded systems. It is widely adopted in avionics, automotive, and IoT~\cite{zou2025automated}, where it supports early architectural analysis, verification, and code generation~\cite{johnsen2012automated,mkaouar2020formal}. 

We evaluate \name on 64 open-source AADL models spanning three RAs: IoT Device System, %
Smart Parking System,  %
and Self-driving Car System. %
We compare \name against three baselines based on lexical~\cite{salton1988termweighting,manning2008introir}, embedding~\cite{reimers2019sentencebert,yin2024study}, and graph~\cite{conte2004thirty} comparison. Results show that, with domain knowledge and GPT-4o, \name achieves 85.8\% overall mapping accuracy, outperforming the strongest baseline by 44.6 percentage points. \name is also the only evaluated approach that enforces the structural constraints of the RA. The presence of domain knowledge increases accuracy by $+8.6$\,pp on average, while adding negligible token overhead.

In a nutshell, this paper makes the following contributions:
\begin{itemize}[leftmargin=*]
\item  We characterize \emph{model-to-RA component mapping} as a constrained matching problem,
\item We propose \name, a novel \emph{LLM-based compliance-guided iterative refinement loop} that corrects RA-level violations, ultimately producing high-quality model-to-RA mappings,
\item We design a \emph{RAG mechanism that combines RA-catalog similarity with similarity over historical mappings} to improve the quality of the discovered mappings,
\item We empirically evaluate \name on 64 AADL models across three RAs against three baselines. All replication artifacts, including code, data, prompts, and ground truth, are publicly available~\cite{GitHubReplication}.
\end{itemize}
\section{Background and Problem Definition}
\label{sec:background}

This section introduces AADL models, rigorously defines the mapping problem with Reference Architectures, and presents a running example used throughout the paper.

\subsection{AADL Models}
\label{sec:bg-aadl}

AADL~\cite{feiler2012model} targets real-time, safety-critical embedded systems and provides specific support for validation and verification~\cite{johnsen2012automated,mkaouar2020formal,feiler2006architecture}. The language combines textual and graphical syntax with execution semantics and uses the XMI interchange format for model persistence and tool interoperability. AADL describes a system as a hierarchical assembly of strongly typed components~\cite{hugues2014analysis}. The left side of Figure~\ref{fig:running-example} shows an abstract visual representation of an AADL model. 

More rigorously, an AADL model  $M = (C_M, E_M)$, where $C_M$ is the set of AADL component instances and $E_M \subseteq C_M \times C_M$ is the set of connectors.
Each component has a \emph{type} that defines its externally visible interface. %
The language distinguishes \emph{software} components (e.g., \texttt{system}, \texttt{process}, \texttt{thread}, \texttt{data}) from \emph{execution platform} components (e.g., \texttt{processor}, \texttt{memory}, \texttt{bus}, \texttt{device}). %
Components interact through \emph{features}, primarily ports with directions (\texttt{in}, \texttt{out}, or \texttt{in/out}). \emph{Connections} link components typically wiring an outgoing port to an incoming port. %

\subsection{Reference Architectures}
\label{sec:bg-ra}

A \emph{Reference Architecture} (RA) is a generic, domain-specific architecture that codifies the structural and functional expectations shared by systems within a class~\cite{angelov2012framework,nakagawa2014reference,cloutier2010concept}. It captures design knowledge validated across multiple projects and supports standardization, reuse, and compliance checking~\cite{garces2021three}. We consider RAs composed of two elements: a textual specification $\textit{RA}_{\textit{spec}}$ and a structural model $\textit{RA}_{struct}$. These two elements are commonly present in many RAs, such as the diagrammatic representations offered by AWS~\cite{aws-architecture}, and Azure~\cite{azure-wellarchitected}.
The \emph{Textual specification $\textit{RA}_{spec}$} is a natural-language description that specifies the intended functional role of the components in the $\textit{RA}_{struct}$. %
The \emph{Structural model $\textit{RA}_{struct}$} represents the essential structure and relationships that characterize systems within a given domain. To represent these models, we refer to the GRAmm metamodel~\cite{bucaioni2023more}, which is a well-known language to represent $\textit{RA}_{struct}$. 
More rigorously, $\textit{RA}_{struct} = (C_{\textit{R}}, H_{\textit{R}}, D_{\textit{R}})$, where:
\begin{itemize}[leftmargin=*]
    \item $C_{\textit{R}}$ is the set of components in the RA. Each component has a set of attributes that include its name and type;
\item $H_R \subseteq C_{\textit{R}}\times C_{\textit{R}}$ is the set of connectors among RA components. Each connector has mandatoriness (mandatory or optional), directionality (unidirectional or bidirectional), and multiplicity attributes;
    
    \item $D_R : C_R \rightarrow \Sigma^*$ associates each RA component with its natural-language description in $\textit{RA}_{\textit{spec}}$.
    
\end{itemize}

\subsection{Problem Definition}
\label{sec:bg-problem}

RAs are used to ensure adherence to established guidelines and principles. Conformance checking is assured by using the concept of incarnation mapping~\cite{rumpe2025tool}, which is a formal specification defining how
elements in a reference model correspond to elements in the concrete model. %

More rigorously, 
a \emph{mapping} from $M = (C_M, E_M)$ to $\textit{RA}_{struct} = (C_{\textit{R}}, H_{\textit{R}}, D_{\textit{R}})$ is a function $\phi: C_M \rightarrow C_{\textit{R}} \cup \{\bot\}$ that assigns to each model component either an RA component or the special symbol $\bot$. The symbol $\bot$ is used when no RA component faithfully represents the role of the AADL component.

The mapping $\phi$, by replacing each component type $c_m$ with the corresponding component type $\phi(c_m)$, generates a version of the model $M$ where concepts are expressed according to the vocabulary of the RA. We refer to this model as $M_{\phi}$.

\smallskip
A mapping $\phi$ is \emph{plausible} for $\textit{RA}_{struct}$ if it is structurally complete and structurally compliant.

\emph{Structurally complete:} There exists no $c_m$ such that $\phi(c_m)=\bot$.

\emph{Structurally compliant.} $M_\phi$ must satisfy the required conformance relation (i.e., incarnation mapping) that enforces the constraints, rules, and characteristics in $RA$. We refer to the definition in~\cite{bucaioni2023more} for a formal definition of the conformance relation. Note that in case engineers want to use a different conformance relation (e.g., one that allows for the violation of a subset of the constraints), any relation can be transparently plugged in \name.

\smallskip

Given a model $M$ and a reference architecture $RA=(\textit{RA}_{struct},\textit{RA}_{spec})$, the model-to-RA mapping problem consists of either discovering that $M$ is not consistent with $\textit{RA}$, or discovering the \emph{correct} mapping between $M$ and $\textit{RA}_{struct}$. Note that a correct mapping is necessarily plausible, but not all plausible mappings are correct. In fact, usually, there exist mappings of the elements in the model into elements of the RA that are logically wrong, that is, they contradict $\textit{RA}_{spec}$. Thus, the challenge is not only finding \emph{a plausible mapping}, but discovering \emph{correct mappings}. 

\subsection{Running Example}
\label{sec:bg-example}

\begin{figure}[t]
  \centering
  \includegraphics[width=\linewidth]{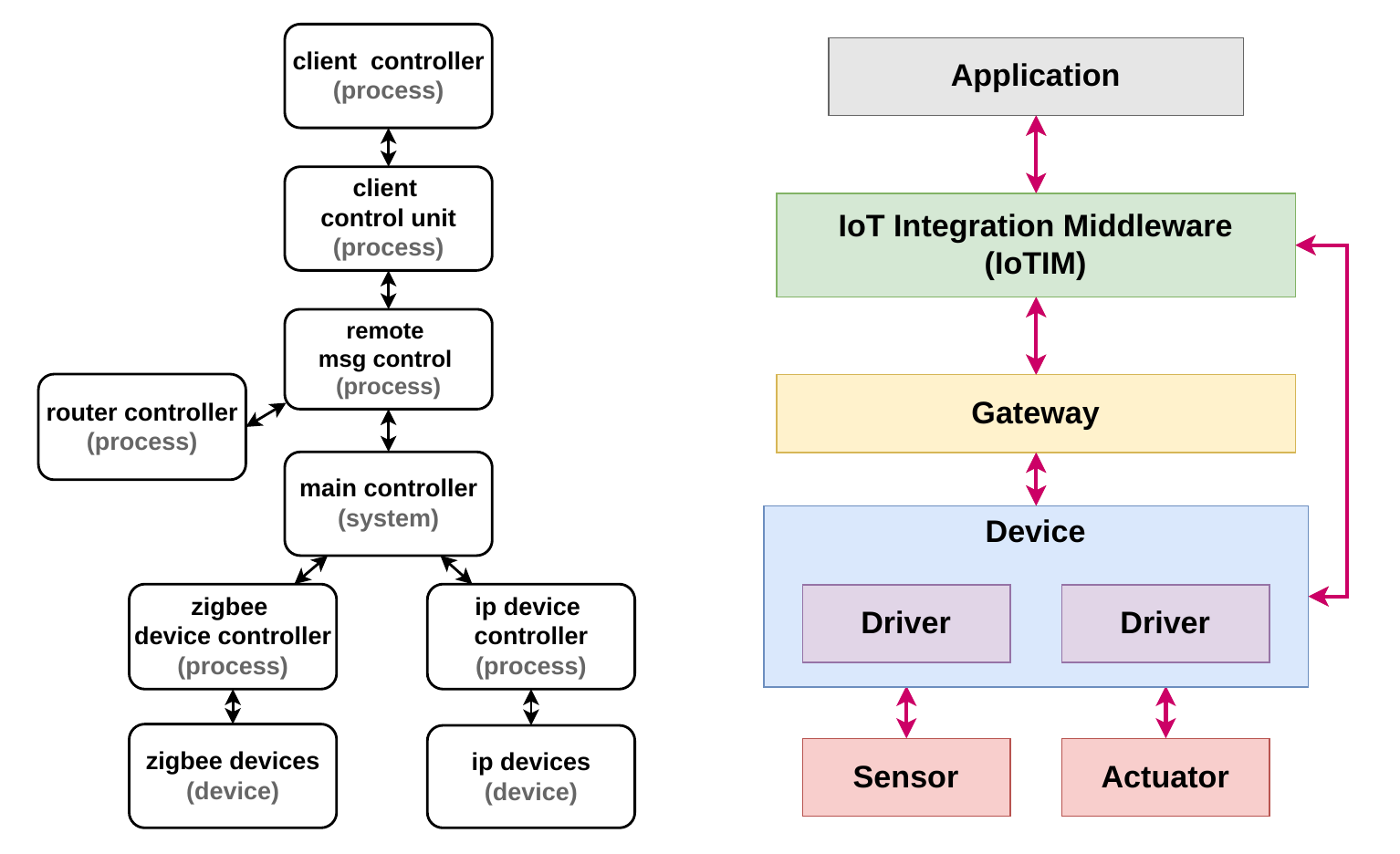}
  \caption{Running example. \textbf{Left:} A Smart Home AADL model with nine components. \textbf{Right:} the IoT Device System RA with seven components. 
  }
  \label{fig:running-example}
\end{figure}

Figure~\ref{fig:running-example} shows a Smart Home AADL model (left) and the IoT Device System RA (right) drawn from the A3D dataset~\cite{dataset}. %
The challenge lies in automatically reconstructing the mapping between them, shown in Figure~\ref{fig:mapping_sts} (c).

The example exposes some of the difficulties in discovering correct mappings. First, the AADL category alone is not sufficient to solve the problem. Six of the nine components share the same category (\texttt{process}), yet they are mapped to three different RA components: \texttt{Application}, \texttt{IoTIM}, and \texttt{Gateway}. The correct assignment requires understanding the functional role of each component, considering %
the AADL component names and the RA descriptions ($\textit{RA}_{\textit{spec}}$ not shown in the figure). A categorical or lexical matcher cannot resolve this gap, as also shown by our empirical results. An effective approach has to reason about each component in its broader context.
Second, the correct assignment must respect the RA conformance relation. The pair \texttt{zigbee\_device\_controller} and \texttt{zigbee\_devices} is connected in the AADL model. Their RA targets must therefore be linked by a relation that the RA permits, either a containment or a connection. Mapping the controller to \texttt{Application} and the device to \texttt{Device} would violate the RA constraints, since \texttt{Application} is neither contained in nor connected to \texttt{Device}. Such structural violations %
emerge only when the candidate mapping is validated as a whole against the RA. This is the role of our compliance check step that drives the iterative refinement loop of our pipeline.
\section{\longName}
\label{sec:methodology}

\begin{figure*}[t]
\centering
\includegraphics[width=0.8\textwidth]{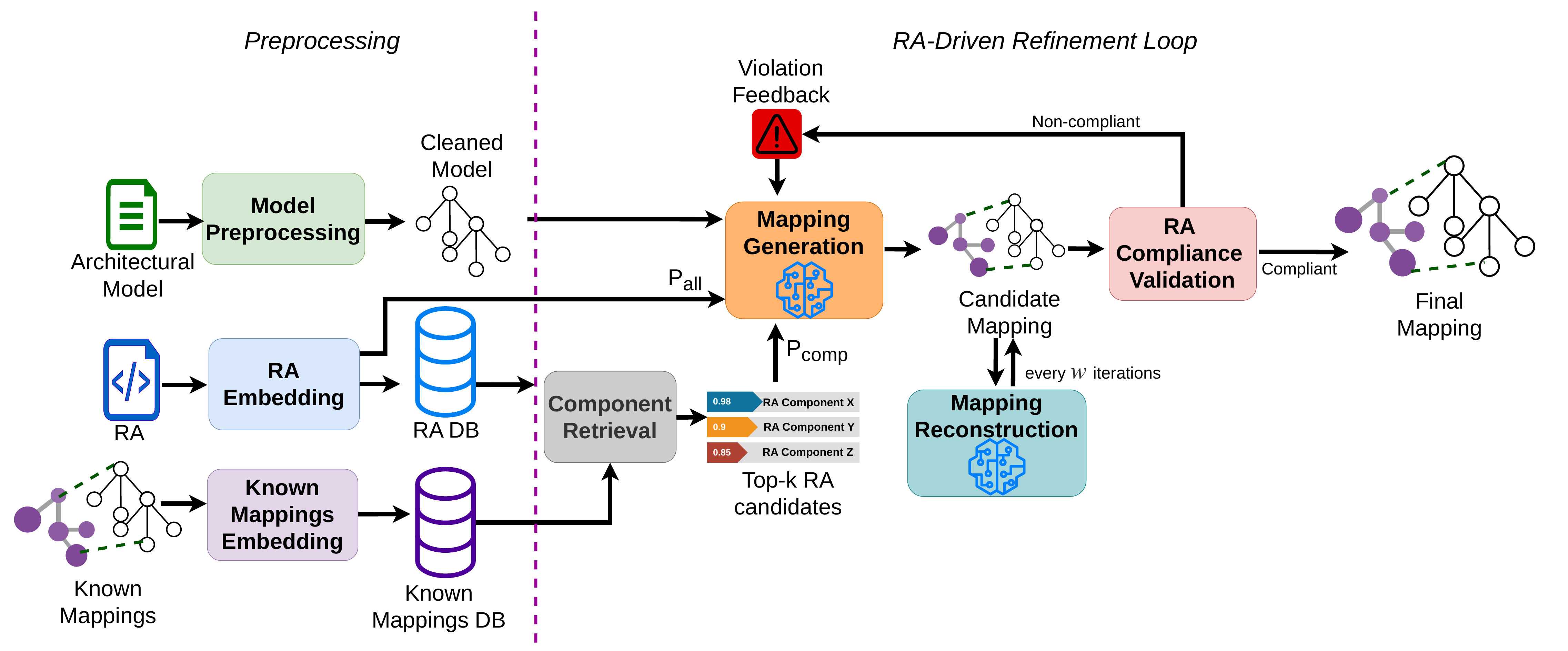}
\caption{Overview of \name. %
}
\label{fig:pipeline}
\end{figure*}

\name takes an AADL model $M = (C_M, E_M)$ and a target reference architecture $RA = (C_R, H_R, D_R)$ as input, and produces a final mapping $\phi : C_M \rightarrow C_R \cup \{\bot\}$ that shows how the model can conform to the RA. Optionally, \name can leverage a set of known mappings $K_{\textit{R}}$ between models and RAs. Figure~\ref{fig:pipeline} illustrates how \name works. 

\name defines a \emph{Preprocessing} stage that prepares the AADL architectural model, the RA, and the known mappings for the analysis, followed by the \emph{RA-driven refinement loop} that iteratively improves the mapping, checking its compliance with the RA until obtaining the final mapping.

The process begins by cleaning the input architectural model (\emph{Model Preprocessing}) to retain only semantically meaningful elements, and by uploading both reference‑architecture components and any available historical domain knowledge into a vector database so they can be semantically retrieved during mapping (\emph{RA and Known Mappings Embedding}). This ensures that the LLM operates on a concise, information‑rich representation of the model and can draw on both RA definitions and prior mappings (\emph{Component Retrieval}).
Using this semantic context, the LLM generates candidate mappings (\emph{Mapping Generation}), either component‑by‑component or for the entire model, and these hypotheses are iteratively refined through compliance validation (\emph{RA Compliance Validation}). To avoid oscillations, the system periodically reconstructs a stable mapping by combining insights from the most recent hypotheses (\emph{Mapping Reconstruction}).

We discuss below how each step is defined.

\subsection{Model Preprocessing}
\label{sec:model-processing}\label{sec:aadl-processing}

The \emph{Model Preprocessing} step removes the irrelevant or potentially noisy details from the model to focus on the elements that carry the most relevant semantic information. 
For this reason, this step extracts the \texttt{name}, \texttt{category}, \texttt{features}, and \texttt{connections} of each component, discarding the rest of the information, such as timing annotations and platform-specific clauses, not contributing to the definition of the role and responsibilities of a component. 

The result, namely the \emph{Cleaned Model}, is serialized into a compact JSON format that is much shorter than the raw model representation, making it more suitable for LLM-based analysis. Listing~\ref{lst:cleaned-model} shows an excerpt of the Cleaned Model for the \texttt{zigbee\_device\_controller} component of the running example.

\begin{lstlisting}[
  language=json,
  caption={Cleaned Model for the \texttt{zigbee\_device\_controller}.},
  label={lst:cleaned-model},
  basicstyle=\ttfamily\footnotesize,
  breaklines=true
]
  "name": "zigbee_device_controller",
  "category": "process",
  "features": [
    {"name": "dev_data_receiver", "dir": "in"},
    {"name": "dev_data_sender",   "dir": "out"},
    ...
  ],
  "connections": [
    "zigbee_devices.out_zd_mcr->zigbee_device_controller.dev_data_receiver",
    ...
  ]
\end{lstlisting}
Model Preprocessing is the only AADL-specific step of \name, since all the following steps operate on the Cleaned Model. Therefore, \name can be adapted to another ADL or modeling notation, such as SysML or UML, by replacing this extractor with one that maps the target notation onto the same JSON schema.

\subsection{RA and Known Mappings Embeddings}
\label{sec:ra-embedding}\label{sec:ra-representation}

To create an RA whose components can be semantically retrieved, for each RA component, \name creates a chunk with the component name, declared features, and description, and embeds it in the RA vector database with a sentence-transformer model~\cite{reimers2019sentencebert}.
Similarly, if the knowledge about some correct mappings between models and RAs is available, \name creates a chunk with every known $(c_m, c_r)$ pair, where $c_m$ is the component that incarnates the component $c_r$ in the RA.  
This chunk is embedded in the Known Mappings vector database and used to inform the mapping procedure. %

\subsection{Mapping Generation}
\label{sec:mapping-generation}

Mapping Generation uses an LLM to analyze the semantic information available and produce a candidate mapping according to two strategies: the \emph{all-at-once strategy} (\Pall), which produces a candidate mapping from the cleaned model all at once; and the  \emph{per-component strategy} (\Pcomp), which produces a candidate mapping by analyzing one component at a time. %

The all-at-once strategy embeds both the whole cleaned model and the full RA description in a single prompt. The models are represented textually by first listing component information, then the edges, and finally the textual descriptions. 

The per‑component strategy maps each component of the cleaned model independently. Each prompt includes information about the component to be mapped together with its direct neighbors to provide the LLM with a minimal degree of structural awareness, even though components are processed one at a time. When no prior knowledge about correct mappings is available, the prompt is simply enriched with the top candidate RA components identified through Component Retrieval (see next section), which searches the RA database for the most semantically similar components to the one under analysis. When historical knowledge is available, Component Retrieval also exploits it to refine the set of retrieved components. Finally, \name aggregates the per-component predictions into a complete mapping that includes all components of the model. %

The all-at-once strategy has the benefit of considering the full model when producing the mapping, thereby leveraging the full contextual information. On the negative side, considering the full model may overwhelm the LLM. The per-component strategy has the benefit of analyzing a component in depth, but the amount of contextual information considered is limited to the neighbors of the analyzed component.

In both cases, each prompt has four blocks.
\begin{itemize}[leftmargin=*]
  \item A \emph{task description} that states the mapping goal and the output that must consist of a list of \textit{(component, RA component)} pairs in a machine-readable format.
  \item The \emph{components to map}. In \Pall, it lists every component that must be mapped. In \Pcomp, it lists a single component and its neighbors.
  \item The \emph{RA context}, namely the relevant RA components, their containment relations, and their descriptions. In \Pall, it includes the full RA. In \Pcomp, the relevant components are the ones returned by Component Retrieval (see next section). 
  \item An optional set of canonical \emph{few-shot examples} drawn from a held-out part of the dataset.
\end{itemize}

Listing~\ref{lst:prompt-excerpt} shows an excerpt of a \Pcomp prompt for the \texttt{zigbee\_device\_controller} component of the running example. The labels in brackets mark the four blocks. The RA descriptions and the few-shot examples are shortened for space. The rest of the prompts used in the evaluation are part of our reproduction package.

\begin{lstlisting}[
  style=promptstyle,
  caption={Excerpt of a \Pcomp prompt for the running example.},
  label={lst:prompt-excerpt}
  ]
[Task description]
Select EXACTLY ONE best label (component name) for the target node. All labels MUST be chosen from the RA component names below.
...

[Components to map]
Target node name: zigbee_device_controller
Neighbor node names:
- zigbee_devices
- main_controller

[RA context]
Reference Architecture component labels (name: description):
- Gateway: Handles communication between devices and higher-level layers when direct connectivity is not possible ...
...

[Few-shot examples]
...
\end{lstlisting}

The mapping generated by the LLM is automatically converted into the XMI format required by the RA compliance checking tool~\cite{bucaioni2023more}. 

When the refinement loop is active, the mapping generation is also executed to improve a non-compliant mapping. In such a case, the prompt changes in two blocks: (i) the task description also includes the list of detected violations and incorporates the recommendation to keep unchanged the mappings that do not cause violations, and to revise only the mappings that cause a violation, and (ii) the structured components to be mapped include the current mapping, and not only a list of components to be mapped. 

\subsection{Component Retrieval}
\label{sec:hybrid-retrieval}

When the per-component strategy is used, the Component Retrieval step automatically retrieves the RA components that must be considered for mapping the given component $c$ from the RA DB and from the Known Mappings DB. From the RA DB, \name retrieves the $k$ semantically closest RA components to $c$ and embeds them in the Mapping Generation prompt. For instance, for the \texttt{zigbee\_device\_controller} component in the running example, the retrieved candidates are \texttt{Gateway}, \texttt{IoTIM}, and \texttt{Device}, as shown in Listing~\ref{lst:prompt-excerpt}.

If the Known Mappings DB is also available, \name first retrieves the $k$ most similar model components to $c$, then reads their ground-truth RA assignments and returns the retrieved RA component as additional candidate suggestions. In particular, \name combines the similarity scores of the items retrieved from the RA db and the Known Mappings db into a single list of RA candidates, and the top $k$ entries are finally selected.

\begin{figure*}[t]
\centering
\includegraphics[width=\textwidth]{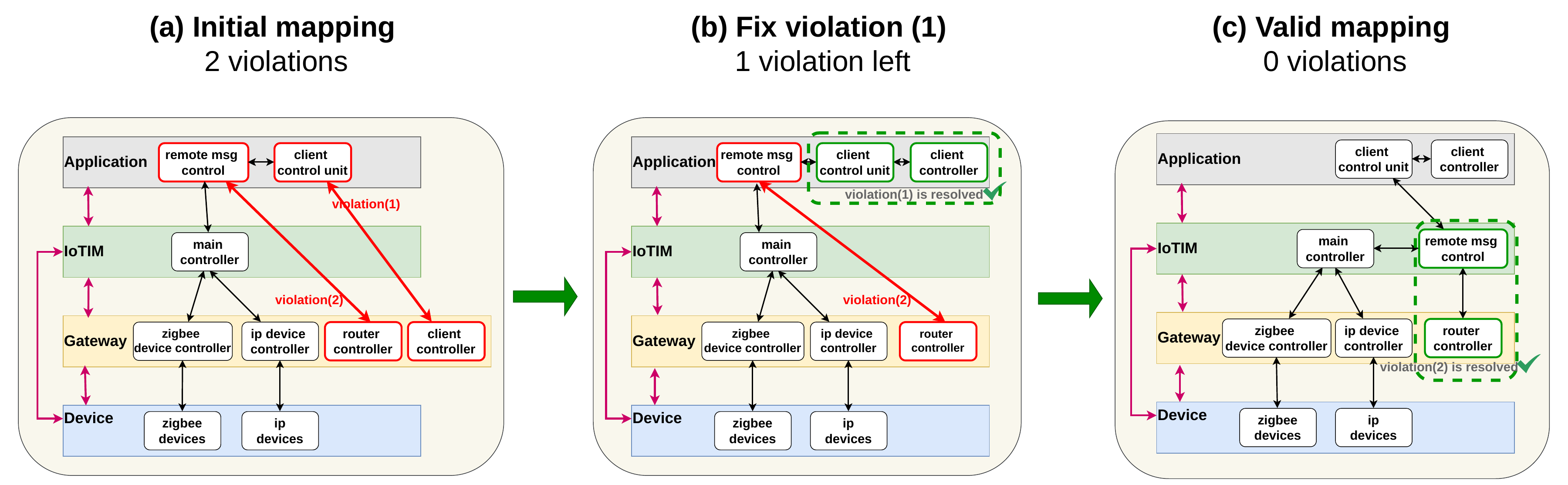}
\caption{Compliance-guided iterative refinement on the running example.}
\label{fig:mapping_sts}
\end{figure*}
\subsection{Iterative RA Compliance-Validation}
\label{sec:loop}

The compliance check decides whether a candidate mapping $\phi$ for a model $M$ is structurally compliant with the available RA. \name uses the rule-based RA validator presented in~\cite{bucaioni2023more} to identify a possibly empty list of violations between $M_\phi$ and $RA$.

If the set of violations is empty, the candidate is compliant, and the loop stops. Otherwise, the list of violations, each consisting of an error message and a list of components causing the error, is used by the Mapping Generation to produce a new candidate solution. 

The loop terminates when one of these conditions holds: (i) the current mapping is compliant with $RA$; (ii) the iteration count reaches the maximum $N_{\max}$; or (iii) the candidate stays unchanged for $p$ consecutive iterations. %

Figure~\ref{fig:mapping_sts} shows how the iterative refinement of a candidate mapping works for the running example. In the first attempt of producing a mapping, \name maps architectural components to RA components as shown in Figure~\ref{fig:mapping_sts}(a). An architectural component $c_m$ drawn inside an RA component $c_r$ indicates that $c_m$ is mapped to $c_r$. In this case, the validator reports two \texttt{relationType} violations. Each one states that the RA allows no connection from Application to Gateway. The validator lists the four offending components, namely \texttt{client control unit}, \texttt{client controller}, \texttt{remote msg control}, and \texttt{router controller}. They appear with a red outline in the figure.
Based on this feedback, the Mapping Generation revises the mapping. It moves \texttt{client controller} from \texttt{Gateway} to \texttt{Application}, shown in the green dashed box. This change removes one of the two forbidden \texttt{Application}-to-\texttt{Gateway} connections. It leaves the compliant assignments untouched, as shown in Figure~\ref{fig:mapping_sts}(b). When checked for violations, the violation count drops to 1. This triggers another refinement round. The Mapping Generation then moves \texttt{remote msg control} to \texttt{IoTIM}, which removes the second forbidden connection. The final result is shown in Figure~\ref{fig:mapping_sts}(c). The final candidate passes all constraints, and \name returns it.

\subsection{Mapping Reconstruction}
\label{sec:reconstruction}

We observed that the computed solution may sometimes oscillate across iterations, with successive candidates trading one violation for another without converging. To break this pattern, \name runs a \emph{reconstruction} step every $w$ iterations. %

The reconstruction prompt receives the last $w$ mapping candidates that have been generated and the corresponding feedback from the validator, and queries the LLM to synthesize a new candidate that merges the correct part of each mapping. The reconstructed mapping replaces the current candidate, and the next iteration starts.

\section{Experimental Setup}
\label{sec:setup}

\subsection{Research Questions}
\label{sec:rqs}

We organize our evaluation around six research questions. %

\begin{description}
\item\textbf{RQ1 (Strategy)} How do the all-at-once and per-component strategies perform?

\item\textbf{RQ2 (Ablation)} How do the individual components of \name affect its effectiveness?%
\item\textbf{RQ3 (Domain Knowledge)} How does the domain knowledge given by the Known Mappings affect the results?
\item\textbf{RQ4 (Architectural Complexity)} How do the characteristics of the AADL models and of the RAs affect the accuracy of \name?
\item\textbf{RQ5 (Comparison with baselines)} How does \name compare to  baseline approaches? %
\item\textbf{RQ6 (Model capacity)} How does the capacity of the underlying LLM influence results?%
\end{description}

\subsection{Dataset}
\label{sec:dataset}

We use a public curated dataset with three RAs of different sizes and domains, paired with a collection of open-source AADL models~\cite{dataset}. To our knowledge, it is currently the only public dataset that pairs architectural models with an explicit RA specification and component-level ground-truth mappings across multiple domains. Other public model collections do not meet this requirement. The benchmark by Fuch{\ss} et al.~\cite{fuchss2022establishing} offers traceability links between architecture documentation and models, while ModelSet~\cite{lopez2024modelset} and the extended EA ModelSet~\cite{glaser2025extended} provide only model-level labels. None of them relates individual components to RA elements, which is the information needed to evaluate a model-to-RA mapping.

Table~\ref{tab:dataset} shows the structure of each RA and the characteristics of the corresponding models used in our experiments. Columns \emph{\#Comp.} and \emph{\#Conn.} report the number of components and connectors~\cite{zhao1998assessing} of each RA and the average number per AADL model. Column \emph{\#Models} reports the number of compliant models for each RA. The last row reports averages, except for \emph{\#Models}, which reports the total. The mappings in the dataset have all been produced after the cutoff dates of the models we used in the evaluation. %

\begin{table}[t]
\centering
\caption{Overview of the dataset: three RAs and the corresponding AADL models.}
\label{tab:dataset}
\renewcommand{\arraystretch}{1.15}
\setlength{\tabcolsep}{3pt}
\begin{tabular}{lccccc}
\toprule
 & \multicolumn{2}{c}{\textbf{RA}} & \multicolumn{3}{c}{\textbf{AADL models}} \\
\cmidrule(lr){2-3}\cmidrule(lr){4-6}
\textbf{RA} & \textbf{\#~Comp.} & \textbf{\#~Conn.} & \textbf{\#~Models} & \textbf{\#~Comp.} & \textbf{\#~Conn.} \\
\midrule
IoT Device       & 7  & 6  & 25 & 7.6  & 7.8 \\
Self-driving Car & 10 & 16 & 10 & 18.1 & 22.4 \\
Smart Parking    & 21 & 22 & 29 & 8.3  & 4.2 \\
\midrule
\textbf{Average}   & 12.7 & 14.7 & \textbf{64}$^{*}$ & 9.6 & 8.4 \\
\bottomrule
\multicolumn{6}{l}{\footnotesize $^{*}$Total number of models.}
\end{tabular}
\end{table}

The RAs span heterogeneous domains. The IoT Device System RA captures the typical decomposition of an IoT deployment, illustrated by the running example of Section~\ref{sec:bg-example}. The Smart Parking System RA models the components of a smart-parking infrastructure, such as parking spots, vehicle detectors, controllers, and user-facing services. The Self-driving Car System RA captures the functional modules of an autonomous vehicle, including perception, localization, planning, control, and the associated sensor and actuator interfaces. All three RAs are provided as Ecore metamodels~\cite{bucaioni2023more,steinberg2008emf} with natural-language descriptions of each component.

The AADL models are open-source instances of systems whose architecture falls into one of the three target domains. The dataset reports the ground-truth mapping between each model and its RA. 

\subsection{Baselines}
\label{sec:baselines}
We considered various classes of techniques as baseline approaches. 

We first considered architecture conformance approaches~\cite{murphy2001reflexion,bucaioni2023more,bucaioni2024continuous,rubei2025violation}. However, they cannot serve as baselines because they all assume the model-to-RA mapping is given rather than discovered. 

We also examined the 67 matching systems named or cited in multiple relevant survey papers~\cite{shvaiko2011ontology,conte2004thirty,shvaiko2005survey,otero2015ontology,bellahsene2010evaluating}, but only 10 of them still provide publicly available source code. Among these, AgreementMaker~\cite{cruz2009agreementmaker} and LogMap~\cite{jimenez2011logmap} are the closest to our setting. However, both operate on OWL and RDF ontologies and rely mainly on lexical matching. As a result, they cannot represent AADL component categories, port directions, or the distinction between containment and connection. Adapting them would require substantial extensions that would bias the comparison. 

To answer RQ5, we instead used as baselines those reference algorithms underlying these matching systems that are applicable to the AADL-to-RA mapping problem. In particular, we identified three baselines operating at the lexical, semantic, and structural level. All the baselines are included in our replication package~\cite{GitHubReplication}. We describe the three selected baselines below.

\paragraph*{TF-IDF lexical matching}
It computes the TF-IDF representation~\cite{salton1988termweighting,manning2008introir} of each AADL component (using its name, features, and category) and of each RA component (using its name and textual description), and associates AADL components with RA components based on their cosine similarity. This baseline captures lexical overlap without semantic generalization.

\paragraph*{Sentence-embedding similarity}
It projects each component into a 384-dimensional semantic space using the pre-trained \texttt{all-MiniLM-L6-v2}~\cite{yin2024study} model. The cosine similarity is again used to establish the mapping between components. This baseline captures semantic overlap.

\paragraph*{Graph-based structural similarity}
This baseline computes mappings by comparing the architectural characteristics of AADL components and RA elements~\cite{conte2004thirty}. Each component is represented by structural attributes, including its type, hierarchical parent, neighboring connections, communication interfaces, and connectivity degree. The similarity between components guides the creation of the mapping.

\subsection{Evaluation Metrics}
\label{sec:metrics}

To answer RQ1-RQ3, RQ5, and RQ6, we use three metrics to characterize the cost-quality trade-offs of the evaluated approaches.

\paragraph{Accuracy}
We adopt the standard accuracy metric to measure the quality of the mapping, defined as the proportion of AADL components on which the produced mapping agrees with the ground truth. For each RA, we compute the accuracy averaged across all AADL models. For the overall comparison, we report average values across the three RAs. 

\paragraph{Runtime}
Runtime is the wall-clock time, in seconds, required to produce the final mapping for a given AADL model. It includes all LLM calls, retrieval operations, and compliance-checking steps. We report the average over the AADL models of each RA.

\paragraph{LLM cost}
LLM cost is the total number of input and output tokens consumed by the LLM during the mapping of a single AADL model. The cost includes the tokens introduced by iterative refinement and mapping reconstruction.

\subsection{Experimental Methodology}
\label{sec:methodology-eval}

\paragraph*{RQ1 (Strategy)}
This RQ compares the all-at-once and per-component strategies described in Section~\ref{sec:mapping-generation}. To directly measure the effect of the strategies, we measure the \emph{quality of the mapping as outputted by these two strategies} without any iterative refinement or mapping reconstruction.   

\paragraph*{RQ2 (Ablation)}
In this research question, we evaluate \name considering \emph{all its components} enabled, with the exception of the Known Mappings, which represent an external domain knowledge that is not always available, and it is thus assessed independently in RQ3. To study the impact of each component of the solution, we also assess \name by selectively removing one element at a time. In particular, for \emph{compliance check}, we run \name with and without the iterative loop presented in Section~\ref{sec:loop}; for \emph{few-shot learning}, we compare zero-shot prompting against a setting with two basic examples covering the case of a process-to-functional and a device-to-physical mapping; %
for \emph{iteration stability}, we compare \name with and without the Mapping Reconstruction step; 
and for \emph{RA representation}, we compare \name using the RA DB fed only with RA component names or fed also with RA component descriptions.

\paragraph*{RQ3 (Domain Knowledge)}
This RQ investigates the impact of the \emph{Component Retrieval} step when domain knowledge is available. The domain knowledge is a set of known mappings stored in the Known Mappings DB. In our experiment, we upload the ground-truth mappings in our dataset, excluding every time the ones about the model under analysis.
We compare \name with and without access to this DB.

\paragraph*{RQ4 (Architectural Complexity)}
This RQ investigates whether the complexity of the AADL models and of the RAs affects the accuracy of \name. For each AADL model, we measure four complexity dimensions, namely size (number of components), connectivity (graph density), hierarchy (maximum containment depth in the AADL instance model), and naming ambiguity (mean entropy of the TF-IDF similarity between each component and the RA components). We compute the Spearman correlation between each dimension and the accuracy on each AADL model, applying the Holm-Bonferroni correction~\cite{holm1979simple} across the four dimensions. For the RAs, we relate accuracy to their number of components and connections.

\paragraph*{RQ5 (Comparison with baselines)}
This RQ compares our approach to the three baselines presented in Section~\ref{sec:baselines}. %

\paragraph*{RQ6 (Model Capacity)}
This RQ investigates the impact of the size of the \emph{LLM} on the results. It compares the results obtained with GPT-4o-mini, which is the model used in every other RQ, to the ones obtained with GPT-4o, to assess the possible benefit of a larger model of the same family.

\smallskip
Since the outputs of LLMs are non-deterministic, for each configuration, we report average values obtained from five repetitions. %
We characterize the magnitude of differences between configurations using Cliff's delta~\cite{cliff2014ordinal}, a non-parametric effect-size measure that is robust to small samples and does not assume a specific distribution. %
Following standard thresholds~\cite{romano2006appropriate}, we interpret $|\delta| < 0.147$ as negligible, $< 0.33$ as small, $< 0.474$ as medium, and $\geq 0.474$ as large. %

\subsection{Tool Implementation and configuration}
\label{sec:impl}

Our \name implementation processes models using the Eclipse Modeling Framework (EMF)~\cite{steinberg2008emf}, and validates them against the RA using the tool implemented in%
~\cite{bucaioni2023more}. All LLM calls go through the official OpenAI API\footnote{\url{https://platform.openai.com/docs/api-reference}} with the temperature fixed to 0 
across all configurations. %
Based on initial runs, we set the maximum iterations $N_{\max} = 30$, same-mapping constraint $p = 3$, and the reconstruction window $w = 10$. In the per-component strategy with hybrid retrieval, the top-$k$ parameter is set to $k = 5$. %
The \name implementation, the prompts used in each configuration, and the raw experimental logs are released on GitHub~\cite{GitHubReplication} and archived on Zenodo~\cite{ZenodoReplication}.
\section{Results and Analysis}
\label{sec:results}

\subsection{\textbf{RQ1}: Strategy}
\label{sec:results-rq1}
We first discuss the accuracy and cost of the \emph{all-at-once strategy} (\Pall) %
and the \emph{per-component strategy} (\Pcomp). %
Table~\ref{tab:PallvsPcomp} shows the results.
\begin{table}[t]
\caption{Comparison between \Pall and \Pcomp across RAs.} \label{tab:PallvsPcomp}
\centering
\resizebox{0.45\textwidth}{!}{%
\begin{tabular}{l|ccc|ccc}
\hline
\multirow{2}{*}{\textbf{RA}}
& \multicolumn{3}{c|}{\textbf{\Pall}} 
& \multicolumn{3}{c}{\textbf{\Pcomp}} \\
& \textbf{acc} & \textbf{tokens} & \textbf{runtime}
& \textbf{acc} & \textbf{tokens} & \textbf{runtime} \\
\hline
IoT  & 66.5\% & 106K  & 564s  & 69.6\% & 1,049K & 5095s \\
Self-driving Car  & 61.2\% & 119K  & 925s  & 63.6\% & 583K   & 3775s \\
Smart Parking     & 59.3\% & 193K  & 1253s & 66.7\% & 802K   & 3930s \\
\hline
\textbf{Overall}  & 62.3\% & 135K  & 868s  & 66.6\% & 789K   & 4228s \\
\hline
\end{tabular}
}
\end{table}
\Pcomp outperforms \Pall by $+4.3$\,pp on average ($+3.1$\,pp IoT, $+2.4$\,pp self-driving, $+7.4$\,pp smart parking), with Cliff's $\delta = +0.40$ (medium). The improvement is consistent across domains. It shows that decomposing a large AADL model into focused per-component queries is more effective than asking the LLM to solve the entire mapping in one shot. The reason is that a single query forces the LLM to reason about all components at once, which dilutes its attention on each mapping decision. \Pcomp instead keeps each query small and centered on one component and its neighbors. However, \Pcomp incurs a 4 to 10 times token overhead and a 3 to 9 times longer runtime, since each component requires a separate API call. %

Ensuring a high-quality and reliable mapping often takes precedence over reducing runtime or token consumption. This is the case especially when better mapping accuracy translates into less effort and time required from architects to interpret and adopt the results. We therefore use \Pcomp for the remainder of the evaluation, with the caveat that \Pall remains a cheaper, albeit less accurate, alternative when needed.

\subsection{\textbf{RQ2}: Ablation}
\label{sec:results-rq2-ablation}
\begin{figure*}[t]
\centering
\includegraphics[width=\textwidth]{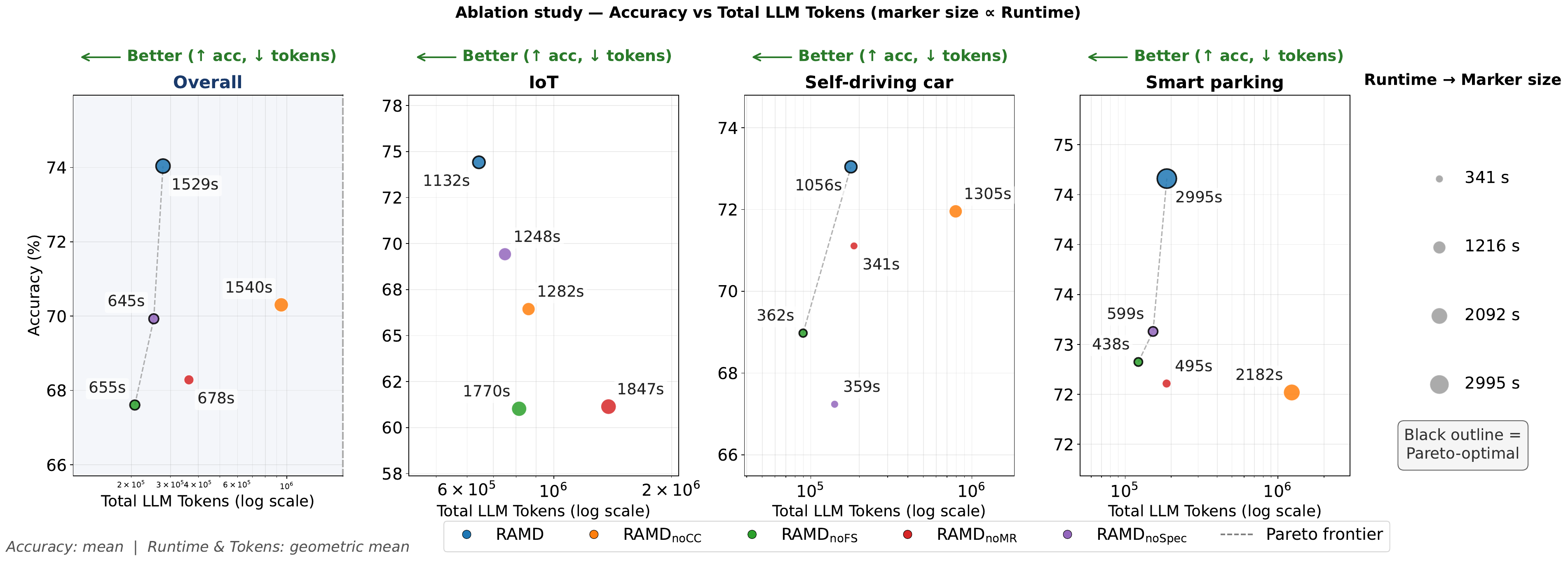}
\caption{Ablation study. Accuracy versus total LLM tokens for each configuration, in each domain and overall. The marker size is proportional to the runtime. The black outline marks the Pareto-optimal configurations.}
\label{fig:rq2-ablation}
\end{figure*}

In this RQ, we measure the effectiveness of \name when the \emph{full workflow is active}, with the only exclusion of the domain knowledge represented by the Known Mappings DB, which is not studied in this RQ but investigated in RQ3. In addition, to assess the impact of each step, we present the results obtained by removing each component: $\name_{noCC}$, which is \name without the conformance checking loop; $\name_{noFS}$, which is \name without the few-shot examples; $\name_{noSpec}$, which is \name using the RA component names instead of the full RA specification $RA_{spec}$; and $\name_{noMR}$, which is \name without the mapping reconstruction step. Figure~\ref{fig:rq2-ablation} reports the accuracy and the token consumption on the two axes. The size of each dot encodes the runtime cost. Configurations on the Pareto front are drawn with a black outline.%

\name reaches the highest accuracy overall, at $74.0\%$, and it is the most accurate configuration in every domain. Each removed component \emph{lowers the accuracy}, which confirms that every step contributes to the final accuracy result. The few-shot examples are the single most important component. Removing them ($\name_{noFS}$) causes the largest accuracy loss, $-6.4$\,pp on average, with the largest effect size, Cliff's $\delta = +0.73$ (large). The mapping reconstruction follows with $-5.7$\,pp (Cliff's $\delta = +0.46$, medium). The RA specification causes a smaller average loss of $-4.1$\,pp, but its effect size is large (Cliff's $\delta = +0.53$), which means that removing the full $RA_{spec}$ produces a difference that is smaller on average yet more consistent across runs. The conformance checking loop has the smallest accuracy impact, $-3.7$\,pp (Cliff's $\delta = +0.42$, medium), but it plays a relevant role on cost and on the guarantee the mapping delivers a technically plausible solution, as discussed below.

Removing a component does not consistently reduce the cost, since its effect varies across domains. A special case is the removal of the conformance checking loop, which causes an increase in cost with $944$K tokens consumed %
and a comparable runtime. Without the compliance feedback, the loop loses its convergence signal and keeps refining until the mapping reconstruction produces a final mapping. The conformance check thus improves accuracy and makes the search more efficient. %

\subsection{\textbf{RQ3}: Domain Knowledge}
\label{sec:results-rq3}
This RQ considers the impact of the domain knowledge stored in the Known Mappings DB (Section~\ref{sec:methodology}).
Table~\ref{tab:RAMDvsRAMDMK} compares \name with RAMD+KM, the variant that retrieves from the Known Mappings DB.

\begin{table}[t]
\caption{Comparison between \name and RAMD+KM across RAs.} \label{tab:RAMDvsRAMDMK}
\centering
    \resizebox{0.45\textwidth}{!}{%
\begin{tabular}{l|ccc|ccc}
\hline
\multirow{2}{*}{\textbf{RA}}
& \multicolumn{3}{c|}{\textbf{\name}} 
& \multicolumn{3}{c}{\textbf{RAMD+KM}} \\
& \textbf{acc} & \textbf{tokens} & \textbf{runtime}
& \textbf{acc} & \textbf{tokens} & \textbf{runtime} \\
\hline
IoT               & 74.4\% & 643K & 1132s & 83.7\% & 868K & 1833s \\
Self-driving Car  & 73.0\% & 178K & 1056s & 82.6\% & 158K & 458s  \\
Smart Parking     & 74.7\% & 187K & 2995s & 81.6\% & 174K & 725s  \\
\hline
\textbf{Overall}  & 74.0\% & 278K & 1529s & 82.6\% & 287K & 848s  \\
\hline
\end{tabular}
}
\end{table}

Domain knowledge has a positive impact on accuracy across all domains. The gain is $+9.3$\,pp on IoT, $+9.6$\,pp on self-driving, and $+6.9$\,pp on smart parking, for an average of $+8.6$\,pp. RAMD+KM reaches an overall accuracy of $82.6\%$, with Cliff's $\delta = +0.91$ (large). Retrieving from the Known Mappings DB provides the LLM with high-quality in-context examples. These examples are selected at runtime for the component under analysis, so the prompt carries evidence of how similar components were mapped in the past.

The token overhead of this gain is small, $1.04\times$ on average. On self-driving and smart parking, the domain knowledge even lowers the token cost to $0.89\times$ and $0.93\times$. The retrieved examples let the compliance loop converge in fewer iterations, which reduces the number of LLM calls. IoT is the only exception. Its larger component vocabulary inflates the retrieval context to $1.35\times$.

The runtime follows the same pattern as the token cost. RAMD+KM is faster on self-driving ($458$s against $1056$s) and on smart parking ($725$s against $2995$s), because the loop converges sooner. The overall runtime drops from $1529$s to $848$s. IoT is again the exception, with the runtime growing from $1132$s to $1833$s, in line with its larger retrieval context.

In a nutshell, domain knowledge significantly improves accuracy. Its effect on cost is small, and in two of the three domains, it is even beneficial.

\subsection{\textbf{RQ4}: Architectural Complexity}
\label{sec:complexity}
\textcolor{black}{Model size is the only dimension with a significant effect. The accuracy of RAMD+KM decreases as AADL models grow larger ($\rho = -0.466$, $p_{adj} < 0.001$), while \name shows the same trend without reaching significance ($\rho = -0.282$, $p_{adj} = 0.095$). In contrast, connectivity, hierarchy, and naming ambiguity show no significant correlation for either approach ($|\rho| \leq 0.22$, $p_{adj} \geq 0.24$). As for the RAs, which are too few to support a statistical analysis, accuracy shows no visible trend with respect to their number of components and connections. } 

In a nutshell, results obtained with large models may require more careful validation than those obtained with smaller ones.

\subsection{\textbf{RQ5}: Comparison with Baselines}
\label{sec:results-rq5}
\begin{figure*}[t]
\centering
\includegraphics[width=\textwidth]{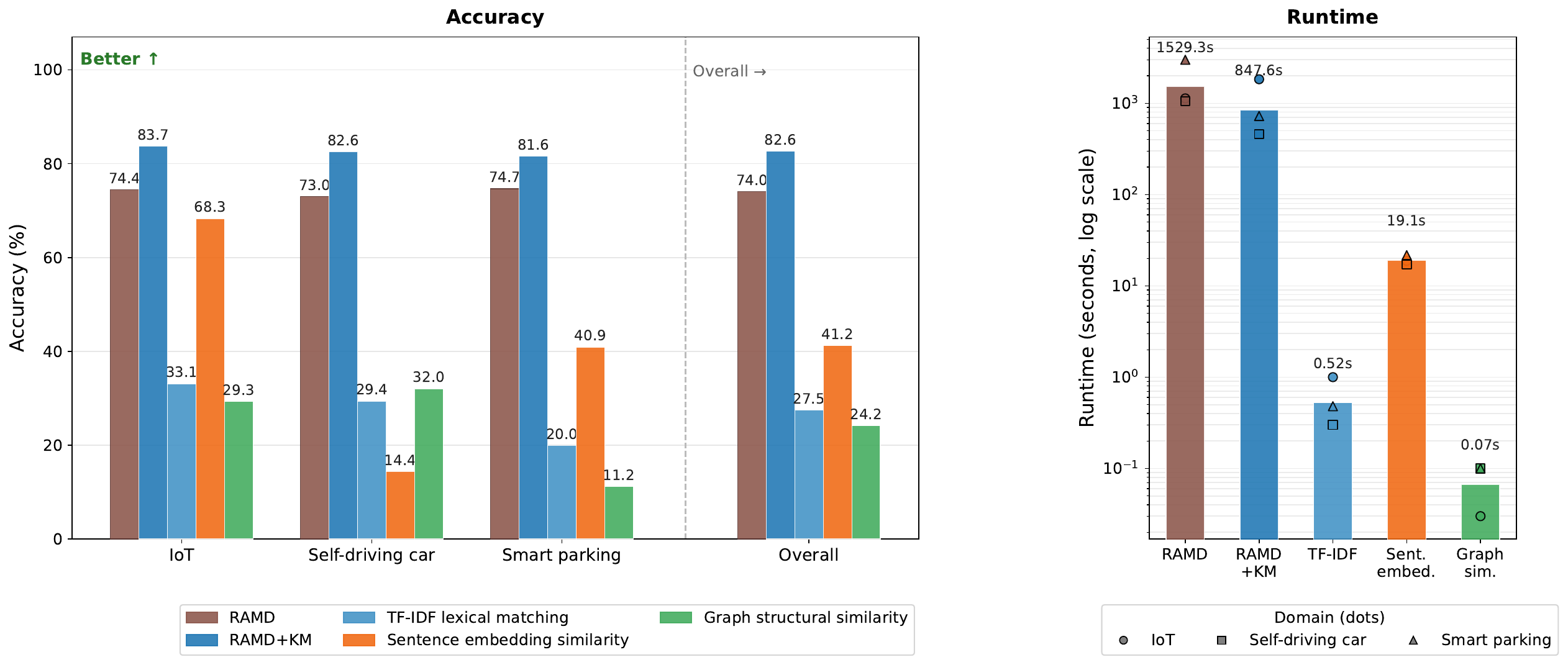}
\caption{Accuracy (\%) per domain and overall (left), and geometric-mean runtime with per-domain points overlaid (right, log scale).}
\label{fig:rq5}
\end{figure*}
Figure~\ref{fig:rq5} compares \name and RAMD+KM to the three baselines.

RAMD+KM outperforms every baseline by a large margin in all three domains. The gap over the strongest baseline of each domain ranges from $+15.4$\,pp to $+50.6$\,pp. On self-driving car and smart parking, no baseline exceeds $40.9\%$, while \name achieves $73.0\%$ accuracy at minimum. The advantage is consistent also without domain knowledge: \name alone overall reaches $74.0\%$ accuracy.    
The effect size is large in every pairwise comparison, with Cliff's $\delta = +1.00$.

The baselines are domain-dependent and often brittle. Sentence-embedding is the best baseline on IoT and smart parking, but it collapses to $14.4\%$ on self-driving, where it fails to capture the domain-specific semantics of that RA. TF-IDF and graph similarity are the weakest methods overall, at $27.5\%$ and $24.2\%$. This confirms that surface lexical overlap and structural heuristics alone are insufficient for reliable mapping.

The accuracy of our approach comes with a higher runtime. RAMD+KM needs $848$\,s on average, while the baselines need at most $19$\,s. The extra time is due to the LLM calls, the RA compliance validation step, and the multiple refinement iterations that identify a high-quality mapping. Given the accuracy gap of $+41.4$\,pp, this trade-off is worthwhile, especially in offline settings without strict timing requirements. Beyond accuracy, our approach is the only one that guarantees conformance with the RA. None of the three baselines can enforce this property.

\subsection{\textbf{RQ6}: Model Capacity}
\label{sec:results-rq6}
This RQ studies the impact of the underlying LLM. We upgrade the model from GPT-4o-mini to GPT-4o, keeping every other component of RAMD+KM
unchanged. Both models belong to the same family, allowing us to isolate the effect of model capacity. Table~\ref{tab:GPT4ominivsGPT4o} reports the results.
\begin{table}[t]
\caption{Comparison between RAMD+KM with GPT-4o-mini and GPT-4o.} 
\centering
\resizebox{0.45\textwidth}{!}{%
\begin{tabular}{l|ccc|ccc}
\hline
\multirow{2}{*}{\textbf{RA}}
& \multicolumn{3}{c|}{\textbf{RAMD+KM (GPT-4o-mini)}} 
& \multicolumn{3}{c}{\textbf{RAMD+KM (GPT-4o)}} \\
& \textbf{acc} & \textbf{tokens} & \textbf{runtime}
& \textbf{acc} & \textbf{tokens} & \textbf{runtime} \\
\hline
IoT               & 83.7\% & 868K & 1833s & 85.2\% & 234K & 470s \\
Self-driving Car  & 82.6\% & 158K & 458s  & 86.8\% & 159K & 275s \\
Smart Parking     & 81.6\% & 174K & 725s  & 85.4\% & 180K & 425s \\
\hline
\textbf{Overall}  & 82.6\% & 287K & 848s  & 85.8\% & 188K & 380s \\
\hline
\end{tabular}
}
\label{tab:GPT4ominivsGPT4o}
\end{table}

The larger model improves the accuracy in every domain. The gain is $+1.5$\,pp on IoT, $+4.2$\,pp on self-driving, and $+3.8$\,pp on smart parking, for an overall gain of $+3.2$\,pp. The peak accuracy is $86.8\%$ on self-driving. GPT-4o reaches an overall accuracy of $85.8\%$, which is the best result of our study and exceeds the strongest baseline by $44.6$\,pp (RQ5). The effect size is large, with Cliff's $\delta = +1.00$.

More notably, GPT-4o also lowers the cost. The mean token count drops to $0.65\times$ of the GPT-4o-mini cost, and the overall runtime drops from $848$s to $380$s. This result may look counter-intuitive, since a larger model is usually more expensive per call. The same effect appears for the compliance loop in RQ2. A more capable model produces correct mappings on the first compliance iteration for most components. This avoids the cascading refinement rounds that drive up the token budget for GPT-4o-mini.

The token saving is uneven across domains. It is largest on IoT, where the count drops to $0.27\times$ ($234$K against $868$K). IoT is the domain where GPT-4o-mini needs the most refinement, as its high token cost in RQ3 shows. A stronger model removes most of these extra iterations, so the saving is concentrated there. On self-driving the token count is almost unchanged ($159$K against $158$K), because GPT-4o-mini already converged quickly on that domain.

In a nutshell, a more capable model raises the overall accuracy to $85.8\%$ while reducing the token and runtime budget.

\subsection{Threats to Validity}
\label{sec:limitations}

\paragraph*{Construct validity}
Construct validity depends on the correctness of the ground-truth mappings. %
We used the ground-truth set publicly released, %
which covers all 64 AADL models in our dataset and was constructed independently of this study. This separation reduces the risk that the evaluation rewards mappings biased toward our approach. %

\paragraph*{Internal validity}
A key factor that may affect internal validity is LLM stochasticity. To address this concern, we set the temperature to 0, made five repeated runs of each configuration, and reported the Cliff's delta to characterize the magnitude of differences.%

\paragraph*{External validity}
External validity is limited by the scale and scope of the evaluation. The dataset covers 64 AADL models across three RAs~\cite{dataset}, which span heterogeneous domains but cannot represent the full range of architectural styles, sizes, and modeling conventions. However, evidence is consistent across three distinct domains, which increases confidence in the results and their validity. 
Furthermore, the evaluation is limited to AADL models. While the approach is largely language-agnostic, its effectiveness on other architecture description languages remains to be investigated.
Finally, replicating the study with LLMs beyond the GPT-4o family, including open-weight ones, is left to future work.
\section{Discussion}
\label{sec:discussion}

This section discusses the main lessons learned from our experiments.

\paragraph*{Domain knowledge can matter more than model capacity}
Known mappings improve accuracy by $8.6$\,pp (RQ3), whereas a larger LLM adds $3.2$\,pp (RQ6). For this reason, building a repository of validated mappings can be a more effective way to improve accuracy than adopting a larger LLM.

\paragraph*{The mapping is a starting point for the architect}
The ground truth contains only correct mappings, so accuracy measures the proportion of components that \name maps correctly, not only plausibly. Even in the best configuration (RQ6), some components receive a mapping that may satisfy the RA constraints but does not match the role described in $\textit{RA}_{spec}$ (Section~\ref{sec:bg-problem}). Consequently, the output of \name still requires manual validation by the architect, especially for large models, whose accuracy is lower (RQ4).

\paragraph*{Violations can reveal incompatible models}
Our dataset only contains AADL models that comply with their RA, but the problem definition also covers incompatible models (Section~\ref{sec:bg-problem}). \name signals this case when the compliance loop does not reach zero violations within $N_{\max}$ iterations (Section~\ref{sec:loop}), or when the reconstructed mapping leaves some components unassigned. To test these signals, we paired three IoT and Smart Home models with the Smart Parking RA. In all three cases, \name returned its best mapping together with unresolved violations involving mandatory components or mandatory connectors. This small experiment gives initial evidence that these violations can reveal an incompatible model.
\section{Related Work}
\label{sec:related}

The problem of relating a concrete artifact to a higher-level reference is studied in two adjacent areas. \emph{Architecture conformance checking} verifies whether an implemented system respects an intended architecture. The reflexion-model technique~\cite{murphy2001reflexion} and subsequent constraint-based and rule-based extensions~\cite{passos2009static,knodel2007comparison,terra2015recommendation} require the architect to declare the intended architecture and to map source-code elements to it, after which the tool detects deviations. These techniques share with ours the goal of relating a concrete artifact to a higher-level reference, but they assume the mapping is available and concentrate on detecting violations. The problem we tackle is the dual one, namely, producing the mapping itself.

\emph{Schema and ontology matching} identifies correspondences between heterogeneous structured artifacts using lexical, embedding-based, and structural similarity signals. Classical surveys and the Euzenat--Shvaiko textbook provide a comprehensive account of the area~\cite{shvaiko2005survey,shvaiko2011ontology,otero2015ontology,bellahsene2010evaluating}. %
These techniques are limited by two factors. First, they treat the two artifacts as sets or generic labeled graphs and do not enforce the structural and mandatory-component constraints that an RA imposes. Second, the heterogeneity gap between a model instance and an RA 
often exceeds the bridging capacity of similarity-based matchers, as confirmed in our empirical evaluation.%

A growing body of work investigates \emph{LLMs in software engineering}~\cite{hou2024large,fan2023large,zhang2026survey} and in model-driven engineering in particular. LLMs have been used to generate or repair Ecore instance models from natural language~\cite{pan2025instance}, to combine model-based specifications with formal verification in the automotive domain~\cite{petrovic2024synergy}, and to translate between informal specifications and domain-specific languages~\cite{di2025use}. A particularly active line of research applies LLMs to \emph{ontology alignment}. He et al.~\cite{he2023exploring} demonstrated the zero-shot capability of GPT and Flan-T5 on OAEI benchmarks. Giglou et al.~\cite{giglou2024llms4om} proposed \textsc{LLMs4OM}, which combines retrieval-augmented generation with zero-shot prompting. Subsequent work has explored complex alignments~\cite{amini2024complexoa}, LLM-based oracles~\cite{lushnei2026large}, and retrieval-driven search strategies~\cite{taboada2025ontology}. These approaches confirm that LLMs are an effective tool for aligning heterogeneous semantic artifacts, but they target unconstrained ontologies rather than RAs. They consequently do not consider the many structural and mandatory-component constraints specific to RAs.

A related line of research is the one on \emph{continuous architectural conformance}~\cite{bucaioni2023more,bucaioni2024continuous,rubei2025violation}. The MORE approach~\cite{bucaioni2023more} models RAs as metamodels and checks the conformance of a Software Architecture to an RA through model-driven techniques. The AssistRA tool~\cite{bucaioni2024continuous} extends this with a continuous-conformance distance function for incremental, non-blocking validation. The work by Rubei et al.~\cite{rubei2025violation} can interact with an LLM-based recommender to suggest a corrected architecture once a violation has been detected. 

\name shares with this line of work the combination of LLM reasoning with formal RA metamodels. %
However, \name addresses a fundamentally different problem. Continuous architectural conformance assumes that the component-to-RA mapping has already been declared by the architect, and focuses on repairing individual violations. \name instead generates compliant model-to-RA mappings without imposing any requirement on the architect. The two classes of approaches can be potentially combined to generate a mapping and then improve it based on the feedback of the architect.

\section{Conclusion}
\label{sec:conclusion}
We presented \name, an approach that discovers the mapping between an architectural model and its RA by integrating model preprocessing, domain knowledge retrieval, LLM-driven semantic analysis, and iterative constraint-checking refinement.
On 64 open-source AADL models across three RAs, \name with domain knowledge and GPT-4o reached $85.8\%$ accuracy, improving over the best baseline by $44.6$\,pp, while meeting the RA structural and mandatory-component constraints by construction. Future work concerns integrating \name with continuous-conformance tools so that its output can serve as the starting point for repairing violations as the model evolves.
Moreover, we will investigate how to exploit the extensive documentation that often accompanies RAs, such as standards and discussions of architectural patterns, to enrich the contextual information available to the LLM.

\bibliographystyle{IEEEtran}
\bibliography{references}

\end{document}